\documentclass[twocolumn]{article}
\usepackage[utf8]{inputenc}

\usepackage[]{algorithm2e, algorithmic}
\usepackage{subfig}
\usepackage{amssymb}
\usepackage{graphicx}
\usepackage{amsmath}
\usepackage{abstract}
\usepackage{bm, float, multicol}
\usepackage{latexsym}
\usepackage{graphicx}
\usepackage{authblk}
\usepackage[left=3cm,right=1.5cm,top=2cm,bottom=2cm,bindingoffset=0cm]{geometry} 
\usepackage[]{amsmath, amssymb}

\SetKwInput{KwInput}{Input}                
\SetKwInput{KwOutput}{Output}
\title{\LARGE \bf Hot-Start Optimization for Variational Quantum Eigensolver}

\author[1]{Belozerova Polina}
\author[2]{Shangareev Arthur}
\author[3]{Zotov Yuriy}
\author[3]{Yung Manhong}
\author[3]{lv Dingshun}
\affil[1]{Moscow Institute if Physics and Technologies, Moscow, Russia}
\affil[2]{Moscow State University, Moscow, Russia}
\affil[3]{Central Research Institute, Huawei Technologies}

\begin{document}
	\thispagestyle{empty}
	\twocolumn[
	\begin{@twocolumnfalse}
		\maketitle
		\begin{abstract}
			The Variational Quantum Eigensolver (VQE) is one the most perspective algorithms for simulation of quantum many body physics that have recently attached a lot of attention and believed would be practical for implementation on the near term quantum devices. However, its feasibility and accuracy critically depend on the ansatz structure, which can be defined in different ways and appropriate choosing the structure presents a bottleneck of the protocol. 
			In our work we present the optimization approach allowing shallow circuit solution. The major achievement of our Hot-Start method is the requiring gates number restriction by several times. In result it increases the fidelity of the final results what is important for the current noise quantum devices implementation. We suggest Hot-Start optimization to be a good new methods beginning for the modern quantum computational devices functionality development.
			\vspace{1cm}
		\end{abstract}
	\end{@twocolumnfalse}
	]

\vspace{1cm}
\section{INTRODUCTION}
	
	One of the main problem of physics and quantum chemistry is to find the many body system ground state. There is no analytical solution for the general case and the computation methods are limited as the complexity of problem in Hilbert space grows exponentially with the number of involved particles. The classical polynomial-time solver does not exist.
	
	Density-functional theory (DFT) \cite{DFT_Engel},\cite{DFT_Sholl} and wavefunction- based methods such as configuration interaction (CI) \cite{CI_method},\cite{QCI} or coupled-cluster (CC) \cite{CC_chemistry}, \cite{CC_computational} are used worldwide for quantum mechanical modeling. For strongly correlated systems tensor network states \cite{CorrSys_tensors} are often used. Each of these methods face exponential computational cost. One of the possible reason is that the number of Slater determinants might be large. Therefore, modern classical computers are limited for tasks, which have applications in new chemical structures search, as well material physical properties study.  
	
	VQE algorithm became the embodiment of Richard Feynman's proposal to simulate one quantum system using another, more controllable \cite{F_cite}. This is possible due to qubits property to form the superposition and keep the entanglement. As a result, the problem is elegantly reduced from the whole ground state $\mathcal{O} (2^n)$ superposition coefficients search to just finding the  $\mathcal{O} (n)$ \cite{Peruzzo2013} angles of qubits vectors rotation and their entanglement.
	
	Despite the significant simplification from an analytical computational point of view, in the practical implementation many difficulties appear. They are related to both the complex goal function surface and the high dimensionality when the number of parameters has to increase.

\vspace{1cm}
\section{QUANTUM CHEMISTRY PROBLEM}

\subsection{Hamiltonian of the System}
In quantum mechanics systems on the atomic scale can be mathematically described by the wavefunction, which accounts all the nuclei and electrons: $\Psi (\textbf{r}_1, \textbf{r}_2, \dots, \textbf{r}_N, \textbf{R}_1, \textbf{R}_2, \dots, \textbf{R}_I )$, where $\textbf{r}_N$ are electrons positions including spin information and $\textbf{R}_I$ are nuclei positions. For condensed matter study time dependence is usually omitted and the wavefunction evolution is described by the Schrödinger equation for many-body Hamiltonian: 

\begin{equation} \label{eq:Schrödinger}
	i \hbar \frac{\partial}{\partial t} \Psi = {\hat{H}}\Psi
\end{equation}

We will use physical constants are set to unity for convenience. The hamiltonian operator is a sum of kinetic energy and potential operators,  $\hat{T}$ and  $ \hat{V}$. The many-body energy operators can be expressed as: 

\begin{equation}
	\hat{H} = \hat{T} + \hat{V}
\end{equation}

\begin{equation} \label{eq:energyT}
	\hat{T} = -\ensuremath{\frac{1}{2}}\left(\sum_N \nabla^2_N + \sum_I \frac{1}{M_I} \nabla^2_I \right) 
\end{equation}

\begin{equation} \label{eq:energyK}
	\hat{V} = \sum_{N,I} \frac{Z_I}{| {\textbf{r}_N - \textbf{R}_I} |} + \sum_{N \leq K } \frac{1}{|\textbf{r}_N - \textbf{r}_K |} + \sum_{I \leq J } \frac{Z_I Z_J}{|{\textbf{R}}_I - {\textbf{R}}_J |}
\end{equation}

Where $ M_I$ are the masses of the nuclei and $\vec{R_I}$ are their coordinates, and $\vec{r}_N$ are coordinates of electrons. The potential energy operator is made up of the Coulomb interactions, neglecting external fields. The three terms are the electron-nuclear, electron-electron and nuclear-nuclear interactions, respectively. The molecular geometry implicitly specification assumes the Born-Oppenheimer approximation with fixed point charges. Thus the ground state electronic energy is a parametric function of their positions. The popular methods to find the wavefunction satisfying the Schrödinger equation, is based on DFT \cite{Grimme1996}.

Thereby the canonical ground state problem may be formulated as an optimization problem with  $\mathcal{O} (2^n)$ complexity, where $\psi$ is a coefficients vector of the resulting superposition in Hilbert space:

\begin{equation}
	\begin{gathered}
		\min \quad \langle \psi | \hat{H} | \psi \rangle\\
		s.t. \quad \| \psi \|_2 = 1
	\end{gathered}
\end{equation}

\subsection{Second quantization form}

After molecule was specified, a particular basis was chosen and the physical anti-symmetry of electrons was enforced, the electronic structure problem may be written exactly in the form of a second quantized electronic Hamiltonian as:

\begin{equation}
	H(\boldsymbol{R}) = \sum_{ij} h_{ij}(\boldsymbol{R}) a^{\dagger}_i a_j + \sum_{ijkl} \frac{1}{2} h_{ijkl}(\boldsymbol{R}) a^{\dagger}_i a^{\dagger}_j a_k a_l    
\end{equation}

Where \textbf{R} is a vector of pairwise distances between particles in the considered system, $a^{\dagger}$ and $a$ are creation and annihilation operators, and $h_{ij}$ and $h_{ijkl}$ are one-electron and two-electron integrals correspondingly \cite{Babbush2016}:

\begin{equation}
	h_{ij} = \int\phi^*_i \left( \textbf{r} \right) \left( -\frac{\nabla^2}{2} - \sum_I \frac{Z_I}{|\textbf{R}_{I} - \textbf{r}|}  \right) \phi_j(\textbf{r}) d \textbf{r}
\end{equation}

\begin{equation}
	h_{ijkl} = \int \frac{\phi^*_i (\textbf{r}_1) \phi^*_j (\textbf{r}_2)    \phi_l (\textbf{r}_1) \phi_k (\textbf{r}_2) }{|\textbf{r}_1 - \textbf{r}_2|} d\textbf{r}_1 d\textbf{r}_2
\end{equation}

For the computational proposes it can be accomplished by OpenFermion and by supported interfaces Psi4 \cite{Parrish2017} and PySCF \cite{Sun2017}. The chosen basis set such as e.g. sto-3g or 6-31g defines the interacting integrals coefficients $h_{ij}$ and $h_{ijkl}$.

\vspace{5mm}
\subsection{Qubit mapping}

After the problem has been written in the second quantized representation, it remains to map it to qubits, as the original problem is formulated for electrons, which are anti-symmetric indistinguishable particles, while qubits are distinguishable particles. There are various of mapping, called encoding, that respect the correct particle statistics. The most popular encodings are the Jordan-Wigner (JW) \cite{JordanWigner1993}, Bravyi-Kitaev (BK)\cite{BravyiKitaev2002} and Bravyi-Kitaev super fast (BKSF) \cite{BravyiKitaev_superfast2018} transformations supported in OpenFermion. In the present work we mostly worked with Bravyi-Kitaev encoding.

\vspace{1cm}
\section{VQE BACKGROUND}

\subsection{Problem reformulation}

The original idea of VQE is the superposition realization on qubits capable to hold a quantum state \cite{Kandala2017}. It restricts the number of optimization parameter to $\mathcal{O} (n)$ angles of qubits rotation and well-tuned entanglement, where $n$ is a number of qubits. Thus, the optimization problem is reformulated to:

\begin{equation}
	\begin{gathered}
	\min \; \langle \psi_{HF} \; U^\dagger (\vec{\theta}) | \; H \; |  \; U(\vec{\theta})  \; \psi_{HF} \rangle = \lambda_{\text{gs}}\\
	s.t. \quad \{{\theta_1}, \dots, {\theta_N} \} \in [ 0, 2\pi ]\\
	\\
	\end{gathered}
\end{equation}

Where $| \psi_{HF} \rangle $ is a Hartree Fock state, the $H$ is problem Hamiltonian and $\lambda_{\text{gs}}$ is a ground state energy. The unitary evolution $U \vec{(\theta)} $ usually is represented as a Pauli strings exponentiation:

\begin{align}
	& U(\vec{\theta}) = U_1 \dots U_N\\ 
	& U_k = U(\theta_k) = \exp{(-i \theta_k P_k)}\\
	& P_k = \mathop{\otimes}\limits_{i=1}^{n} \sigma_{ik}\\
	&\sigma_i \in \{I, X, Y, Z\} \text{  - Pauli matrices}
\end{align}

forming the evolution of trial state to the ground state.

\subsection{The existing methods and the bottleneck}

One of the VQE research direction is to find a proper structure of ansatz for particular Hamiltonian \cite{Grimsley2019TrotterUCCSD}  \cite{Grimsley2019AdaptVQE}. The final circuit has to be comprehensive for optimization in terms of rotation angles and entanglement. As there is no analytical solution for the circuit, this problem remains to be unsolved and to be a wide field for the research.

However, there are two generally accepted methods for quantum chemistry simulations via VQE:  
\begin{itemize}
	\item UCC-based ansatzes \cite{Romero2018} and 
	\item hardware efficient \cite{Kandala2017}.
\end{itemize}
 The UCCSD ansatz is obtained by replacing the traditional Hermitian operator terms in coupled cluster's theory with anti-Hermitian operators. It has a complex structure and it is complicated to implement on quantum hardware due to the large number of gates.  

Hardware efficient ansatzes were proposed for NISQ use. The main idea is to stack similar layers - depths - of rotation and entanglement operators several times \cite{Kandala2017} \cite{SalisMoll2019} \cite{Mitarai2019}. Increasing the number of depths repeatedly one gives the additional chances for the algorithm to converge and in this way the final optimization error has a tend to decrease. The full form of layer circuit is:

\begin{equation}
	\centering
	\begin{gathered}
		\min \; \langle \psi_{HF} \; U^\dagger \vec{(\theta_1)} \dots U^\dagger  \vec{(\theta_d)}  \; | \; H \; |  \; U \vec{(\theta_d)} \dots U \vec{(\theta_1)}  \; \psi_{HF} \rangle\\
		s.t. \quad \{ \vec{\theta_1}, \dots,  \vec{\theta_d} \} \in [ 0, 2\pi ]^{\otimes nd}
	\end{gathered}
\end{equation}

But the computational cost grows as a number of optimization parameters $\mathcal{O}(nd)$. It makes complicated to use VQE for big quantum systems due to optimization methods limitations: 

\begin{itemize}
	\item first order optimization methods require thousands of iteration but
	\item second order methods become unable to deal with number of parameters (angles) due to the large number Jacobian elements and fails because of accumulative computational errors. 
\end{itemize}


Such ambiguity in the number of parameters  in the classical part of VQE forms the bottleneck. Also there are two methods - analytical and computational - to estimate gradient for the optimization but each of them also has disadvantages for the described problem. 

\vspace{2mm}
\textbf{The analytical} method includes the derivative of  subject function:

\begin{align}
	f(\vec{\theta}) = \langle H \rangle_{\vec{\theta}} &=  \langle 0 \; U^\dagger_{\theta_1} \dots U^\dagger_{\theta_D}  \; | \; H \; |  \; U_{\theta_D} \dots U_{\theta_1}  \; 0 \rangle\\
	 \frac{\partial }{\partial \theta_k } U_{\theta_k} &= - i \theta_k P_k \cdot \exp{( - i \theta_k P_k)} = V_{\theta_k} 
\end{align}

Introducing the convenient notation one can derive:

\begin{align}
	 U_{-} &= U_{\theta_{1}} \dots U_{\theta_{k-1}}\\
	 U_{+} &= U_{\theta_{k+1}} \dots U_{\theta_{D}}
\end{align}

\begin{align}
	\frac{\partial }{\partial \theta_k } f(\vec{\theta}) = \frac{\partial }{\partial \theta_k }  \langle H \rangle_{\vec{\theta}} = \langle 0 \; U_{-}^\dagger \; | \; [V_{\theta_k}, \; U_{+}^{\dagger} H U_{+} ] \; | \; U_{-}^\dagger \; 0 \rangle
\end{align}

Thereby, corresponding to Poisson bracket the analytical method requires to estimate expectation value of the quantum scheme twice for each parameter of optimization.

\vspace{2mm}
\textbf{The computational} method is exactly the explicit difference scheme which requires to call expectation function $nd + 1 = D + 1$ times for first order accuracy method:

\begin{align*}
	\frac{\partial f}{\partial \theta_k } = \frac{ f(\theta_1, \dots, \theta_k +\Delta \theta_k, \dots \theta_D) - f(\theta_1, \dots, \theta_k, \dots \theta_D) }{\Delta \theta_k}\\
\end{align*}

If one uses the second-order accuracy scheme, it will appropriately require $2D$ iterations to compute gradient each time. As a result all approaches are clumsy and do not simplify optimization of any order.

Methods mentioned above have one thing in common: they do not take into account any property of the original objective function.

\vspace{1cm}
\section{HOT-START METHOD}

\begin{figure*}[th!]
	\centering
	\subfloat[H2 4 qubits \label{subfig-1:dummy}]{%
		\includegraphics[scale=0.55]{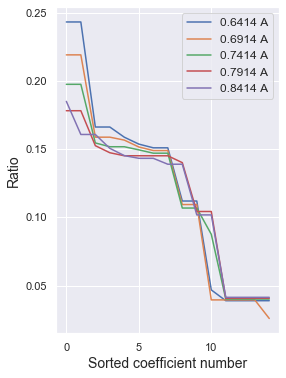}
	}
	\subfloat[LiH 6 qubits version \label{subfig-2:dummy}]{%
		\includegraphics[scale=0.55]{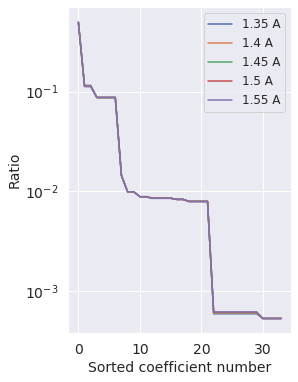}
	}
	\subfloat[LiH 8 qubits version \label{subfig-2:dummy}]{%
		\includegraphics[scale=0.55]{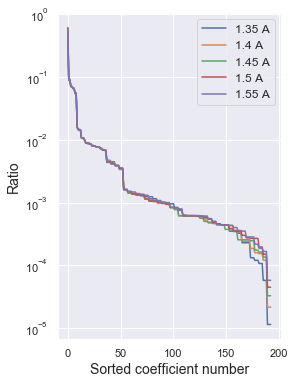}
	}
	
	\caption{\small Maximum possible integral coefficients contribution to the ground state value computed as FCI enegy. For all molecules sto-g3 orbital basis was used. (a) Ordinary 4 qubits H2. (b) Restricted 6 qubits LiH version was received by [0, 1, 2] electrons freezing. (c) 8 qubits LiH has [0, 4] frozen electrons}
	\label{fig:coefs_contributios}
\end{figure*}

\subsection{New Approach Idea}

Previously not an optimization methods but ansatzes were build basing on the main property of Hamiltonian and its Pauli strings based structure:

\begin{align}
	& H = \sum_{k} h_k P_k, \quad P_k = \mathop{\otimes}\limits_{i=1}^{n} \sigma_{ik}\\
	& |U_k| = |\exp(-i\theta_k P_k )| = 1
\end{align}	

But in our work we focused on the analogous property of Pauli strings - eigenvalues $\lambda $ modulo value:

\begin{align}
	\lambda (P) = \pm 1
\end{align}

This property might be interpreted as a maximum possible Hamiltonian sum member contribution to the target ground state.

In one of the last works \cite{GarciaSaez2018} authors demonstrated the adiabatic method using a linear approximation of the problem: $H(t) = A(t) H_0 + B(t) H $. In this case all Pauli strings will contribute to the resulting expectation value simultaneously and continuously over time. It makes the convergence dependence on Hamiltonian structure complex, because in adiabatic approach one adds \textit{all} Pauli strings at the same time uncontrolled. But in comparison the elementary parts  contribution is very clear and equal to $\langle \psi |P_k| \psi \rangle $. 

In contrast, Hot-Start method proposed in this work approximates the objective function step by step, complicating its structure gradually. Our idea is the discrete approach with limited function representation iterations. It is based on the set of Pauli strings and corresponding integral coefficients. Hamiltonian construction by the algorithm allows to manage matrix sparsity slowly.

First, we studied the absolute values distribution for the near ground state Hamiltonians for $H_2$ and $LiH$ molecules. As it can be seen from the Figure \ref{fig:coefs_contributios} the received distributions are highly inhomogeneous. The largest absolute integral coefficient tends to the ground state energy value. The analytical explanation this prevalent effect is that Number and Coulomb/exchange operators contribute to the ground state more than correlation terms as Excitation, Number-excitation and Double excitation operators, which was mentioned in \cite{BravyiKitaev_superfast2018}. These operators define the significance of the integral influence to the ground state.

One may notice that in sorted absolute values of Hamiltonian coefficients the received distribution first members slope is especially sharp. This means that the largest elements may have the great contribution to the ground state value. We propose that solving the problem for the largest elements in the first place may help to avoid local minimum caused by the rest part of the Hamiltonian. The fastest sorting algorithm has $\mathcal{O}(K \log K)$ complexity, where max $K$ is a Pauli string basis size equal to $4^n$. This is a high complexity, but as in practice the number of Hamiltonian sum is much less wherein computational cost is not so resource demanding for many systems of interest.





After the sorting is successfully done the smooth approximation of the objective function to the final one demonstrated meaningful results. The physical interpretation of this approach allows us to call this method a Hot-Start - we gradually rotate the qubits to their best angles, focusing first on the most important Hamiltonian elements, starting from point supposedly close to the absolute minimum. Thus Hot-Start is a method based on the Hamiltonian properties, and this is the objective function based optimization.

\subsection{The algorithm}

Hot-Start method is represented in \textbf{Algorithm 1} for encoded Hamiltonian.

\begin{algorithm}[h!]
	\label{alg:the_alg}
	\KwInput{Starting parameters are fixed random angles of initial gates rotations $\vec{\theta}$}
	\KwOutput{Ground state of $H$}
	\KwData{\\Encoded $H = \sum_i^N h_i \mathop{\otimes}\limits_{j=1}^{n} \sigma_{ij} =  \sum_i^N H_i  $  of the given system, where $\sigma_{j} \in \{I, X, Y, Z \}$
		}
	\quad \\
	\begin{itemize}
		\item Sort $H_i$ in the descending order of the $h_i$ \\ 
		absolute values.
		\item Receive the sequence of \{ $\hat{H}_i$ \}, where $|\hat{h}_1| \geq |\hat{h}_2| \geq \dots \geq |\hat{h}_N| $. 
		\item Start iterative optimization:
	\end{itemize}
	\quad \\
	$k = 1$\\
	\While{ $ k \leq N$ }{
		$A_k = \sum_{1}^k \hat{H}_i$ \\
		Solve by chosen method:\\
		\quad \\
		$\vec{\theta_{k+1}} = \text{argmin} \; \langle \psi \;  U^\dagger (\vec{\theta_k}) \;  | \; A_k \; | \;  U (\vec{\theta_k}) \;  \psi  \rangle $\\
		k = k +1}
	\KwResult{$\vec{\theta_{N+1}}$ is a final set of angles for the original VQE problem}
	\caption{Hot-Start method}
	\vspace{7mm}
\end{algorithm}

\begin{figure}[th!]
	 \label{subfig-1:orders_log}
	 \includegraphics[width=0.48\textwidth]{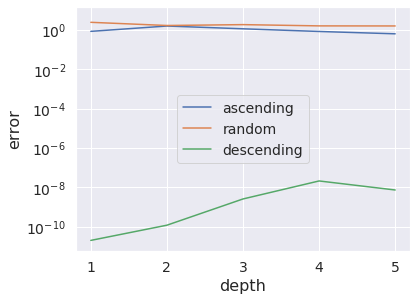}
	 \caption{
	 	\small Different approaches Hamiltonian approximation convergence error comparison for 6-qubits model with 34 Pauli strings. The BFGS optimization method was used.}
	 \label{fig:orders_compare}
\end{figure}

Since sorting part of the algorithm may have high asymptotic complexity firstly we studied the importance of members selection method: we additionally conducted tests with Hamiltonian elements sorted in ascending order and also added randomly. The results are shown in the Figure ~\ref{fig:orders_compare}.

\vspace{1mm}

\vspace{1cm}
\section{RESULTS}

\begin{figure*}[th!]
	\centering
	\subfloat[H2 4 qubits \label{subfig-1:4qubits }]{%
		\includegraphics[scale=0.45]{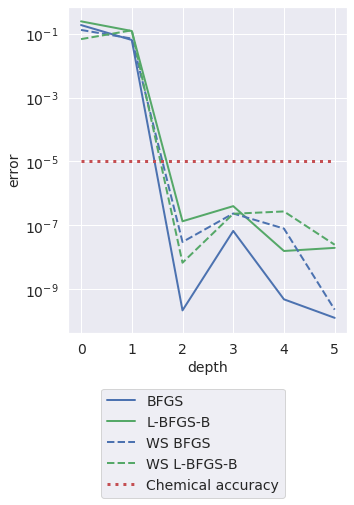}
	}
	\subfloat[LiH 6 qubits \label{subfig-2:6qubits}]{%
		\includegraphics[scale=0.45]{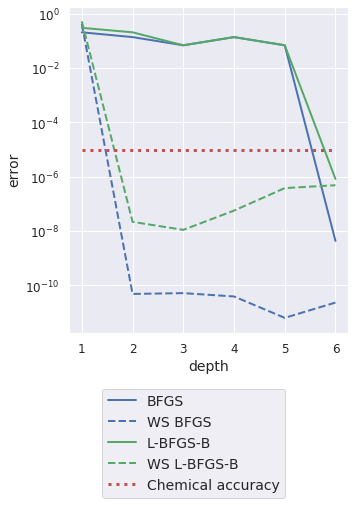}
	}
	\subfloat[LiH 8 qubits \label{subfig-2:8qubits}]{%
		\includegraphics[scale=0.45]{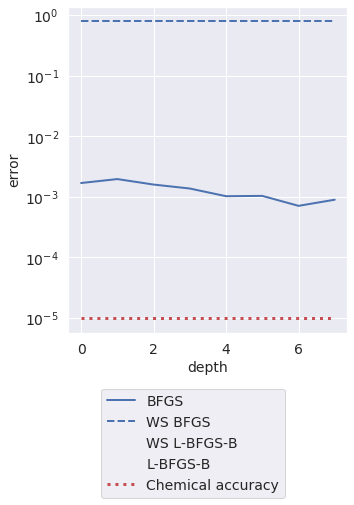}
	}
	
	\caption{\small Average VQE convergence error for Branyi-Kitaev encoding and Ry ansatzes. For 6-qubits model LiH with frozen [0, 1, 2] orbitals in sto-3g basis was used in ~\ref{subfig-1:4qubits }}
	\label{fig:resuts_WS_vs_ordinary}
\end{figure*}

The second order algorithms such as BFGS and L-BFGS-B were used due to their high efficiency to test the ordinary approach and Hot-Start optimization. To research whether Hamiltonian-based optimization can work better than the Hamiltonian-based ansatz we compare the average error of convergence to the known ground state. Corresponding ground state values were computed by eigen-decomposition as number of qubits could be run efficiently on simulator for chemistry problem is not high. It was possible to test the algorithm only on a narrow range of qubits such as:

\begin{itemize}
	\item 4 qubits H2, sto-3g basis;
	\item 6 qubits restricted LiH, sto-3g basis;
	\item 8 qubits restricted LiH, sto-3g basis.
\end{itemize}

The reduction of LiH to receive 6-qubits model was done by freezing [0, 3, 4] orbitals as it corresponds to chemical configuration priority. This orbital combination produces 118 Pauli strings. As Hot-Start method has a high sensitivity to the number of Hamiltonian elements, hence we also designed artificial problems with less number of Pauli strings:

\begin{itemize}
	\item 6-qubits LiH with frozen [0, 1, 2] orbitals and Hamiltonian with 34 sum members;
	\item 8 qubits LiH with frozen [0, 3] orbitals and Hamiltonian with chosen only 10 elements - each 20th member of it
\end{itemize}

Bravyi-Kitaev encoding, Ry ansatz and CNOT full-entanglement gates were used as quantum circuits parameters. The optimization was conducted by 10 randomizations to estimate the uniform convergence ability of methods.

The results with desired chemical accuracy of convergence are represented in the Figure ~\ref{fig:resuts_WS_vs_ordinary}. We would interpret results for 4 qubits in a way that the original problem is simple enough by itself and that is why it was not a high difference for ordinary and Hot-Start optimization. 

For 6 qubits LiH model with 34 Pauli strings we found that Hot-Start optimization works much better than usual approach: the ground state of the simple BFGS method error remains to be about $10^{-1}$ Hartree while Hot-Start approach converges well to $10^{-10}$ Hartree beating the chemical accuracy at the very shallow depth. For the ordinary approach at least 6 depths are required to converge when Hot-Start optimization requires three times less - only 2 depths.

Results for 8 qubits model demonstrate, that for the ground state is hardly achievable for any kind of optimization method.

Thereby there is a precedent case when convergence is achievable even for shallow circuit. And despite the Hot-Start method can not be widely universal due to its linear dependence on number of Pauli strings, it demonstrates the ability to optimize problem with less number of gates. This is important observation for quantum hardware implementation.

\vspace{1cm}
\section{DISCUSSION AND CONCLUSIONS}


We have presented series of results to demonstrate the possibility of Hamiltonian based optimization approaches. The Hot-Start method is the most direct one and might be computationally hard but it proved an important idea: if ordinary optimization methods do not converge to the ground state well, the problem might be not in the chosen ansatz, but in the optimization method for the object function overall.

The represented Hot-Start results can be a good beginning for such a type of optimization, when one does not need anymore to use UCCSD ansatzes to converge, because this type of Hamiltonian-based circuits is large, noisy and complicated for physical implementation. But a new methods require a guess about the optimization process itself. We would propose that may be there is an optimal batch optimization exists, where the batch is a group of Hamiltonian Pauli strings united with some general property.

The most obvious and intuitive would be expect physical features to work well. We divided the Hamiltonian into groups as: Number, Coulomb/exchange, Excitation, Number-excitation and Double excitation operators. This approach has an advantage that in this way Hamiltonian has a fixed number of groups - five - and complexity does not depend on the $\mathcal{O}(n^{4})$ possible Pauli strings.

We implemented the realization of these experiments but revealed that it does not work well for described molecules and other cases. Accordingly to these results the formulation of the batch optimization remains to be an open question.




\addtolength{\textheight}{-2cm}   


\end{document}